# Magnetic phase transitions in SmCoAsO

V.P.S. Awana[*,1], I Nowik[2], Anand Pal[1], K. Yamaura[3,4], E. Takayama-Muromachi[3,4,5] and I. Felner[2]

[1]National Physical Laboratory (CSIR), Dr. K.S. Krishnan Marg, New Delhi-110012, India

[2]Racah Institute of Physics, The Hebrew University, Jerusalem, 91904, Israel

[3]JST, Transformative Research-Project on Iron Pnictides (TRIP), Tsukuba, Ibaraki 305-0044, Japan

[4]Superconducting Materials Center, National Institute for Materials Science, 1-1 Namiki, Tsukuba, Ibaraki 305-0044, Japan

[5]International Center for Materials Nanoarchitectonics (MANA), National Institute for Materials Science, Tsukuba, Ibaraki 305-0044, Japan

Magnetization, x-ray diffraction and specific-heat measurements reveal that SmCoAsO undergoes three magnetic phase transitions. A ferromagnetic transition attributed to the Co ions, emerges at $T_C$=57 K with a small saturation moment of ~0.15 $\mu_B$/Co. Reorientation of the Co moment to an antiferromagnetic state is obtained at $T_{N2}$=45 K. The relative high paramagnetic effective moment $P_{eff}$=1.57 $\mu_B$/Co indicates an itinerant ferromagnetic state of the Co sublattice. The third magnetic transition at $T_{N1}$=5 K is observed clearly in the specific-heat study only. Both magnetic and $^{57}$Fe Mossbauer studies show that substitution of small quantities of Fe for Co was unsuccessful.

[*] Corresponding Author

Dr. V.P.S. Awana

National Physical Laboratory, Dr. K.S. Krishnan Marg (CSIR), New Delhi-110012, India

Fax No. 0091-11-45609310: Phone No. 0091-11-45608329

e-mail-awana@mail.nplindia.ernet.in: www.freewebs.com/vpsawana/

# Introduction

The recent discovery of superconductivity (SC) at relative high temperatures in F doped LaFeAsO (denoted as the 1111 family) has stimulated a large number of experimental and theoretical studies of the materials containing Fe-As layers as a structural unit [1]. In short period of time other Fe-As layers based superconducting families such as: $BaFe_2As_2$ (122) and LiFeAs have been discovered. The optimal doping for superconductivity in 1111 system was studied by: (i) F substitution on the O sites, (ii) by inducing oxygen deficiency, or (iii) by partial substitution of the trivalent Rare-earth (RE) ions by bi-or tetravalent cationic species. All these efforts led to a considerable increase of $T_C$ up to 55 K [1].

One of the most intriguing features of this class of materials is related to their magnetic properties. Similar to the high Tc superconductors (HTSC), in the parent compounds, such as (REFeAsO) a structural distortion, which usually precedes a long range antiferromagnetic (AFM) ordering, is observed at ~150K. After doping, the AFM ordering is suppressed and the compounds become SC and exhibit a good metallic behavior down to $T_C$. Whether magnetic fluctuations are involved in the development of SC is an open question that could give new insight on the analogous problem related to HTSC.

A step further, is to substitute completely Co for Fe. In the pure LaCoAsO, small moments on Co order ferromagnetically (FM) below ~ 60 K with saturation moment ($M_S$) of 0.3-0.4 $\mu_B$/Co [2, 3]. Above the transition temperature, the magnetic susceptibility exhibits a large temperature dependence which follows the Curie-Weiss law. The relative high paramagnetic (PM) magnetic moment deduced is $P_{eff}$~1.3 $\mu_B$, suggests that LaCoAsO is an itinerant ferromagnet and that its magnetic properties are governed by spin fluctuations [4]. Recent neutron diffraction study reveals that the isostructural NdCoAsO undergoes three magnetic transitions. At 69 K, the Co ions become ferromagnetically ordered with a small $M_S$ ~0.2 $\mu_B$. The second transition at 14 K is related to an antiferromagnetic transition of the Co ions with propagation vector (0 0 1/2) and a $M_s$ of ~0.4 $\mu_B$ The third magnetic transition at 3.5 K with a larger moment of 1.3 $\mu_B$ is attributed to the Nd ions which orders AFM with the same propagation vector [5]. When Fe is partially substituted with small amount of Co, SC is induced and the mixed $SmFe_{1-x}Co_xAsO$ materials are SC with $T_C$ =14 K and 15.5 K for x= 0.10 and 0.15 respectively [6].

Here we report our results on a polycrystalline SmCoAsO material which exhibits similar magnetic properties as the two isostructural materials described above. Magnetic measurements performed at low dc applied fields (10-20 Oe) show two magnetic transitions at 57 K and 45 K, which are related to FM and AFM transitions respectively. A third magnetic transition around 5 K related to Sm (not detected directly by magnetic studies) is clearly observed by specific heat study. By replacing Co with a small quantity of Fe, multi-phase materials are obtained. $^{57}$Fe Mossbauer measurements show that the major fraction of Fe ions reside in the Co site, but the magnetic features of the mixed $SmCo_{1-x}Fe_xAsO$ (x=0.05-0.2) samples are masked by the presence of magnetic extra phase(s) which exist in the samples.

## Experimental Details

Polycrystalline SmCoAsO and Fe doped samples were synthesized by a single step solid-state reaction method. Stoichiometric amounts of Sm, Fe, As, and $Co_3O_4$ powders were ground thoroughly under high purity argon atmosphere. The mixed powders were palletized, vacuum-sealed ($10^{-4}$ Torr) in a quartz tube and then subsequently heated (i) at 550$^{o}$C for 12 hours (ii) at 850$^{o}$C for additional 12 hours and then (iii) at 1150$^{o}$C for 33 hours. The furnace was turning off and cooled down to ambient temperature. The crystal structure was studied by powder X-ray diffraction (XRD), using Rigaku diffractometer with Cu $K_{\alpha}$ radiation. Temperature dependence dc magnetization curves at various applied fields were performed in a commercial MPMS5 Quantum Design SQUID magnetometer. Heat capacity studies were carried out by a Quantum Design PPMS system. Mössbauer studies were performed using a conventional constant acceleration drive and a 50 mCi $^{57}$Co:Rh source. The isomer shifts (I.S.) values reported are relative to that of iron**.**

## Results and Discussion

Due to the similarity of the ionic radii of $Fe^{2+}$ (0.75Å) and $Co^{2+}$(0.79Å) in the low spin mode, the lattice parameters of SmCoAsO are very similar to that of SmFeAsO. Room-temperature XRD pattern for SmCoAsO exhibited in Fig. 1, shows that all diffraction peaks can be assigned to the tetragonal crystal phase (space group of *P*4/*nmm*), ensuring the purity of the studied sample. The lattice parameters obtained are: *a*=3.957(3) Å and *c*=8.242 (2) Å. These are somewhat higher (*a*-parameter) and smaller (*c*-parameter) than *a*=3.937(3) *c*=8.498 (2) Å reported for SmFeAsO [1, 6]. The unit cell volume of SmCoAsO is slightly smaller than as for SmFeAsO, which is in agreement with an earlier report [7]. The Rietveld refinement of the XRD pattern of SmCoAsO, yields the structural parameters listed in Table 1.

The temperature dependence of the zero-field-cooled (ZFC) and field-cooled (FC) curves for SmCoAsO measured at H= 20 Oe is depicted in Fig. 2. The peaks at $T_C$=57 K and $T_{N2}$=45 K are attributed to FM and AFM transitions of the Co sublattice, respectively. In that sense this magnetic behavior is similar to that reported for NdCoAsO [5]. The two magnetic transitions are sensitive to the applied external field and shift to lower temperatures for higher H values. It appears that $T_{N2}$=45 K for the AFM transition is much higher than 14 K measured for NdCoAsO. The possible reasoning behind the same may be the difference in magnetic moments and more importantly the varying inter atomic distances between (Nd/Sm)-Co across c-direction [7]. The slight deviation at low temperatures of the two ZFC and FC curves is related to the AFM state of Sm sublattice to be discussed later.

The isothermal magnetizations at various temperatures have been measured up to 10 kOe. In Fig. 3 we present three representative plots: (i) at 15 K, 50 K and at 75 K which are well below, in the middle and above the two magnetic transitions shown in (Fig. 2). These curves support the two types of magnetic transitions described above. The linear dependence of M(H) at 15 K is a conclusive evidence for the AFM order existing below 45 K. On the other hand a typical FM M(H) plot is obtained at 50 K. At high H values the magnetization is almost saturated and the saturation moment of Co, deduced by extrapolation of this curve to H=0 yields $M_S$ ~0.15 $\mu_B$/Co, a value which is very close to $M_S$ ~0.2 $\mu_B$ obtained by a neutron diffraction study of of NdCoAsO in its FM state. At 75K a linear M(H) is observed indicating a normal paramagnetic (PM) state. The dc susceptibility (measured at 20 Oe) $\chi$(T) (=M(T)/H) at elevated temperatures, adheres closely by the Curie-Weiss (CW) law: $\chi = \chi_0 + C/(T-\theta)$, where $\chi_0$ is the temperature independent part of $\chi$, C is the Curie constant, and $\theta$ is the CW temperature. The values extracted by a least square fit (not shown) are: C= 0.42(1) emu T/mol Oe and a positive $\theta$= 68.7(1) K which is consistent with the FM state at $T_C$ of SmFeAsO. The highest expected PM moment for $Sm^{3+}$ is 0.85 $\mu_B$. Subtracting this value from C yields $P_{eff}$ =1.57 $\mu_B$/Co, a value which is close enough to $P_{eff}$ =1.3 $\mu_B$/Co reported for LaCoAsO [4]. The low $M_S$ obtained and the high $P_{eff}$ /$M_S$ ratio both suggest that SmCoAsO is also an itinerant FM as discussed in details in Ref. 4.

The zero-field temperature dependence of the specific-heat ($C_P$) of SmCoAsO is shown in Fig. 4. For the sake of comparison $C_P$ of SmFeAsO [8] is also shown. It is readily observed that basically the two curve overlap each other in the entire temperature range, except for the distinct kink around 140 K observed for SmFeAsO, which is related to its spin density wave magnetic anomaly. The sharp peak at $T_{N1}$ ~ 5 K for both compounds is due to the AFM ordering of $Sm^{3+}$ sublattice, indicating the same C for the two substances. Note that the areas below the two peaks are very close to each other. It is possible that the absence of the AFM transition in the magnetization curves (Fig. 2) is due to the relative small moment of $Sm^{3+}$ which is masked by the Co ions magnetization. Thus, we may assume that the slight deviation of the two ZFC and FC curves above 5 K (Fig. 2) is due to some magnetic fluctuations of the Sm moments induced by the Co moments.

Substitution of Fe in $SmCo_{1-x}Fe_xAsO$: In a previous publication it was shown that for in SmFeAsO, partially substitution of Fe by small Co content, is successful and nearly single phase materials are obtained. Moreover, SC is induced and for the mixed $SmFe_{1-x}Co_xAsO$ materials $T_C$ = 15.5 K for x= 0.15 is achieved [6]. Using the same token, in $SmCo_{1-x}Fe_xAsO$ we partially substituted Fe for Co up to x=0.25. XRD studies (not shown) show clearly the existence of unidentified extra phases even for x as low as 0.05 and that their amount increases with x. Figs 5 and 6 illustrate this observation.

Fig. 5 shows the ZFC and FC magnetization curves of $SmCo_{0.8}Fe_{0.2}AsO$, which both exhibit a typical FM state behavior up to 221 K. This FM state arises from a magnetic impurity phase. $^{57}$Fe Mossbauer studies (MS) have been performed on various $SmCo_{1-x}Fe_xAsO$ compound at 300 and 90 K. Generally speaking, all spectra obtained were similar to each other. Fig. 6 exhibits the MS of $SmCo_{0.8}Fe_{0.2}AsO$ at 90 K which is below the FM magnetic transition. The major line corresponds to divalent Fe ions which reside in the original Co *(2b)* crystallographic site. Least square fit yields the hyperfine parameters: I.S. = 0.50(1) mm/s, a quadrupole splitting of 0.179(1) mm/s and a line width of 0.29 mm/s. These values fit well with $^{57}$Fe MS already published for the 1111 family [9]. The relative intensity of the magnetic sextet is 22% with parameters: I.S.=0.91 mm/s, quadrupole shift =-0.25 mm/s and a magnetic hyperfine field of $H_{eff}$=187 kOe, corresponds to an unknown magnetic phase ( probably to some Sm-Co-Fe based phase) as stated above. We may conclude that, in contrast to the parent SmFeAsO compound in which small quantities of Co can be introduced, no single phase materials can be obtained in the $SmCo_{1-x}Fe_xAsO$ system.

# Conclusions

We have shown that SmFeAsO undergoes three magnetic transitions at low temperatures. Specific heat studies reveal a peak $T_{N1}$ ~ 5 K which is attributed AFM transition of the $Sm^{3+}$ sublattice. A bulk FM phase transition of the Co ions is observed at $T_C$= 57 K. The extrapolated saturation moment in the FM state is ~0.15 $\mu_B$. On the other hand, the PM effective moment is $P_{eff}$ =1.57 $\mu_B$, indicating an itinerant FM for the Co sublattice. A third magnetic AFM transition at $T_{N=}$ 45 is attributed to reorientation of the Co moment into an AFM magnetic state. In that sense the qualitative magnetic behavior of SmFeAsO is very similar to that of NdCoAsO. However, the magnetic behavior of SmFeAsO differs from that reported for NdFeAsO in a quantitative sense, because of the influence of the larger magnetic moment of Sm in ordering Co. Also varying inter atomic distances between (Nd/Sm)-Co across c-direction [7] may influence the resultant magnetic behavior of SmCoAsO and NdCoAsO. These results are meaningful in the sense that they highlight the fact that RE ions moments in these compounds influence the 3d metal ordering in adjacent layer (Co-O) in the structure. In contrast to $SmFe_{1-x}Co_xAsO$ x<0.25, XRD, magnetic and $^{57}Fe$ Mossbauer studies show that partially substitution of Fe for Co in $SmCo_{1-x}Fe_xAsO$ (even for x=0.05), leads to formation of extra magnetic phases.

Acknowledgments: This research is partially supported by the Israel Science Foundation (ISF, Bikura 459/09), and by the Klachky Foundation for Superconductivity. This work is partly supported by Indo-Japan DST-JSPS bilateral research program and MANA visiting scientist scheme of NIMS Japan. Anand Pal acknowledges the financial support from CSIR in terms of CSIR-NET Fellowship. Dr. H. Kishan (HOD) and Prof. R.C. Budhani (Director-NPL) are acknowledged by VPSA and Anand Pal for their encouragement.

Table 1 Rietveld refined structure parameters of SmCoAsO, [$R_p$ = 3.85, $R_{wp}$= 4.93, $R_{exp}$= 3.22, $\chi^2$= 2.34 ]

| *Atom* | *site* | *x* | *y* | *z* | *Occupancy* |
|---|---|---|---|---|---|
| Sm | *2c* | 0.25 | 0.25 | 0.1356 | 1.00 |
| Co | *2b* | 0.75 | 0.25 | 0.5000 | 1.00 |
| As | *2c* | 0.25 | 0.25 | 0.6514 | 1.00 |
| O | *2a* | 0.75 | 0.25 | 0.0000 | 1.00 |

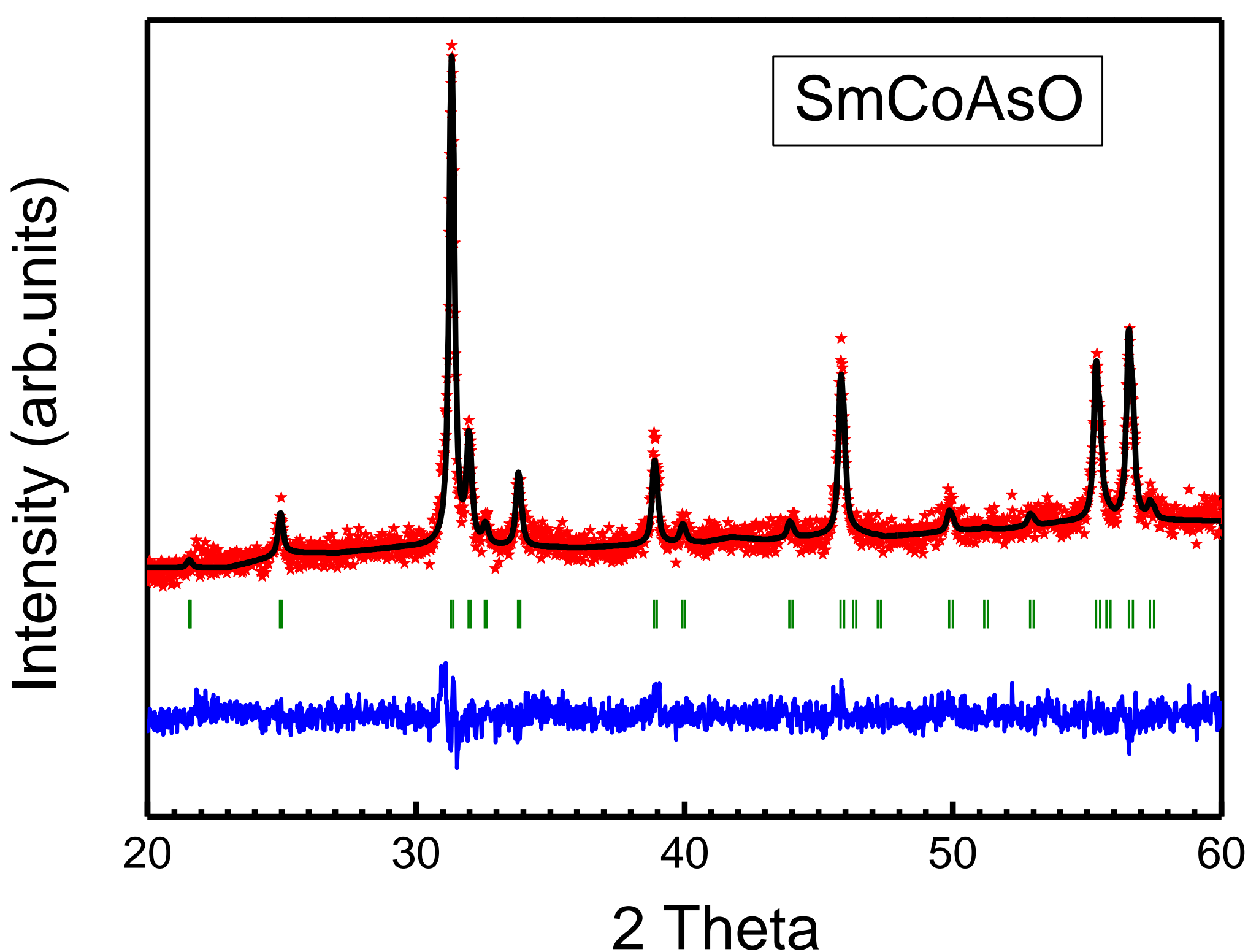

Figure 1: A room temperature X-ray diffraction pattren of SmCoAsO.

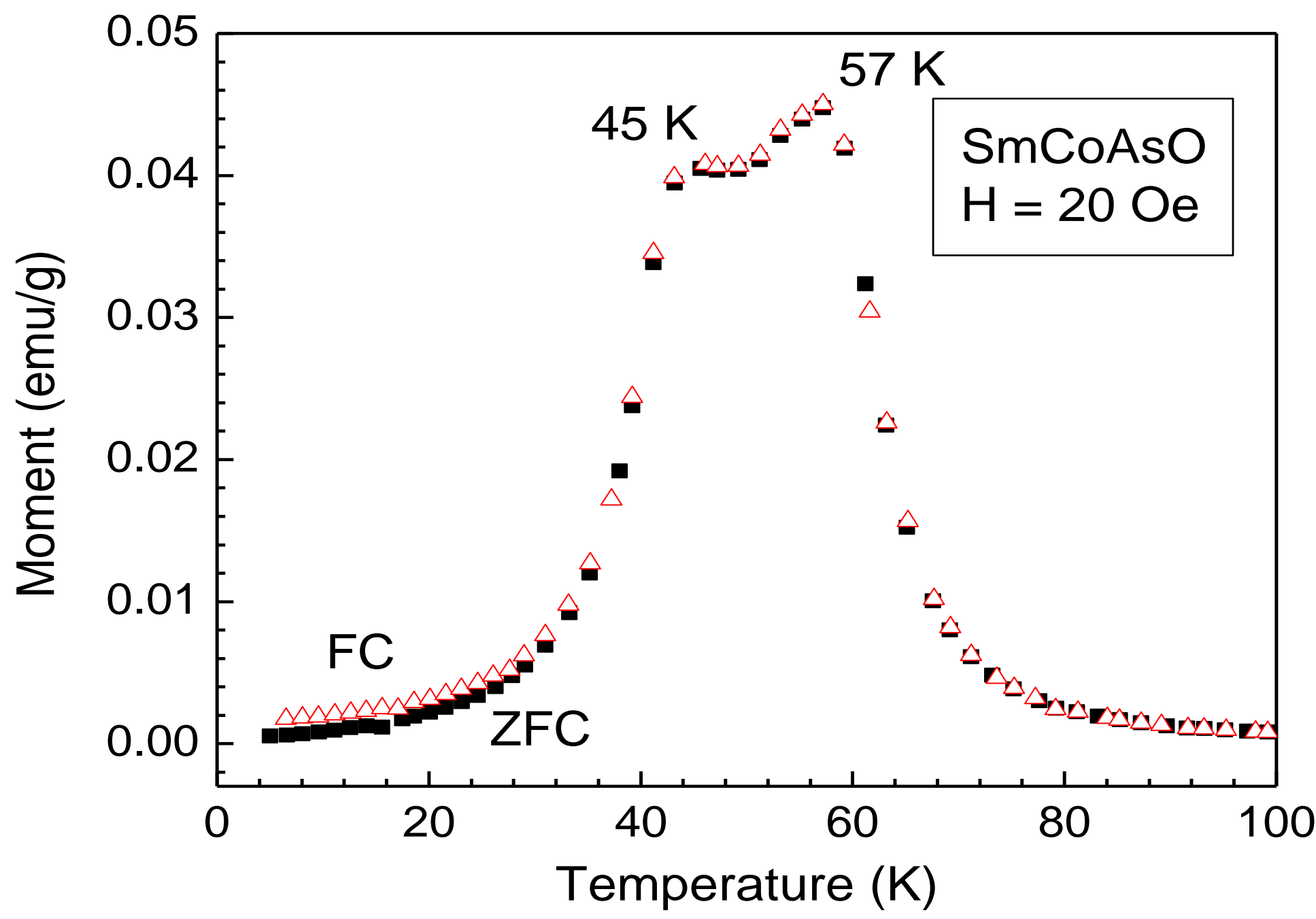

Figure 2: ZFC and FC plots of SmCoAsO measured at 20 Oe.

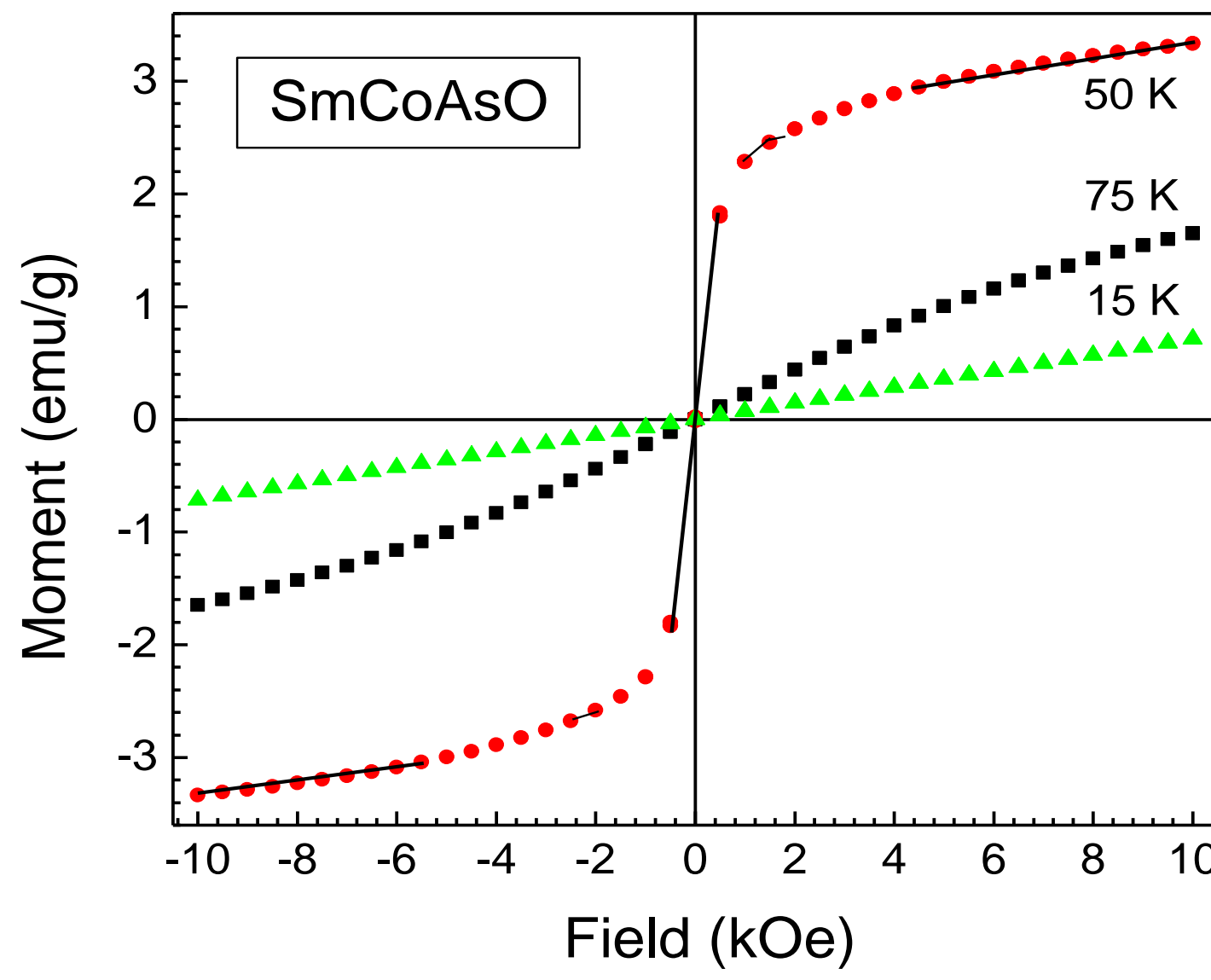

Figure 3: Selected isothermal magnetization curves of SmCoAsO at typical temperatures. Note the FM like curve measured at 50 K.

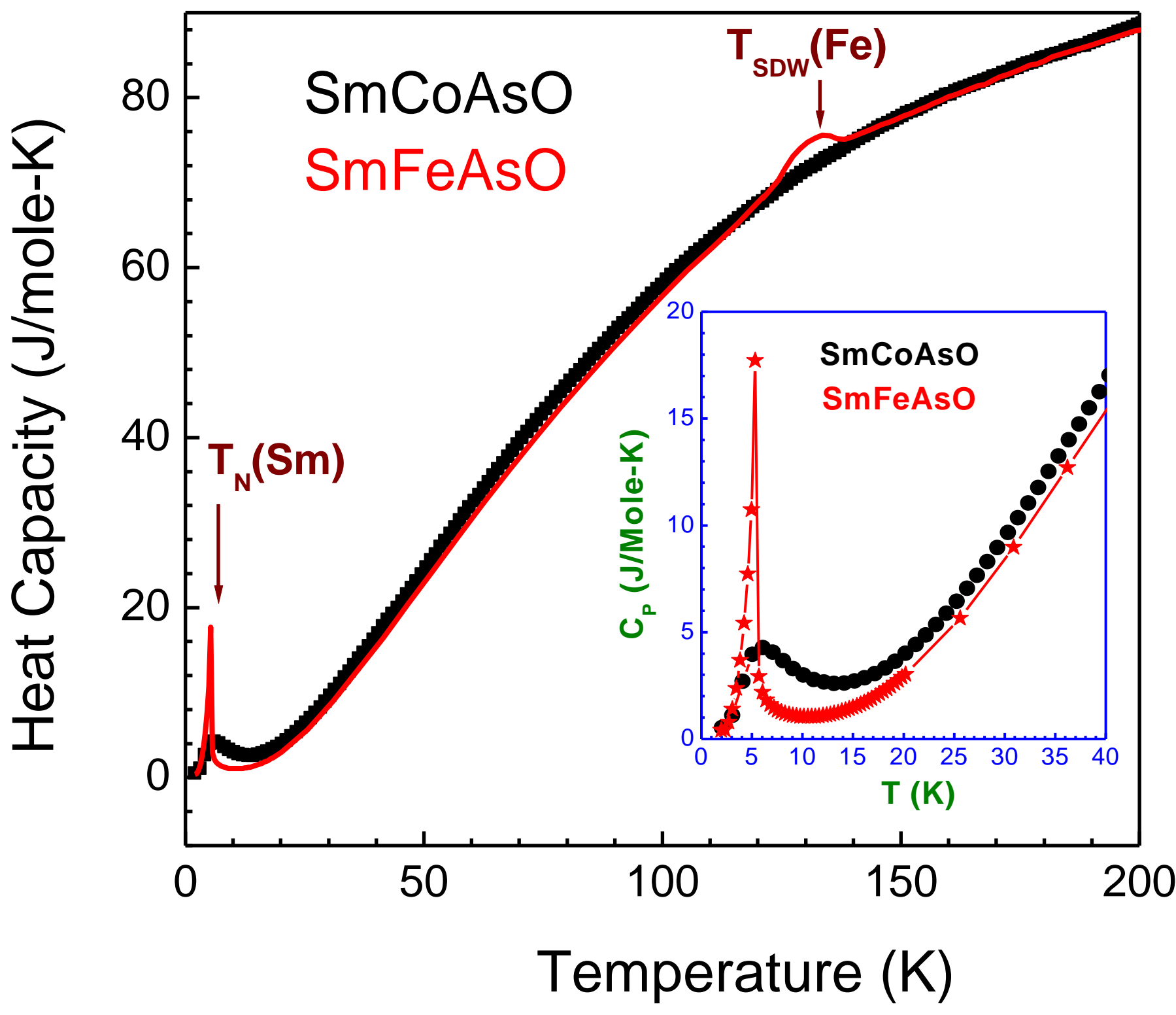

Figure 4: Zero-field specific heat curves of SmCoAsO and SmFeAsO samples.

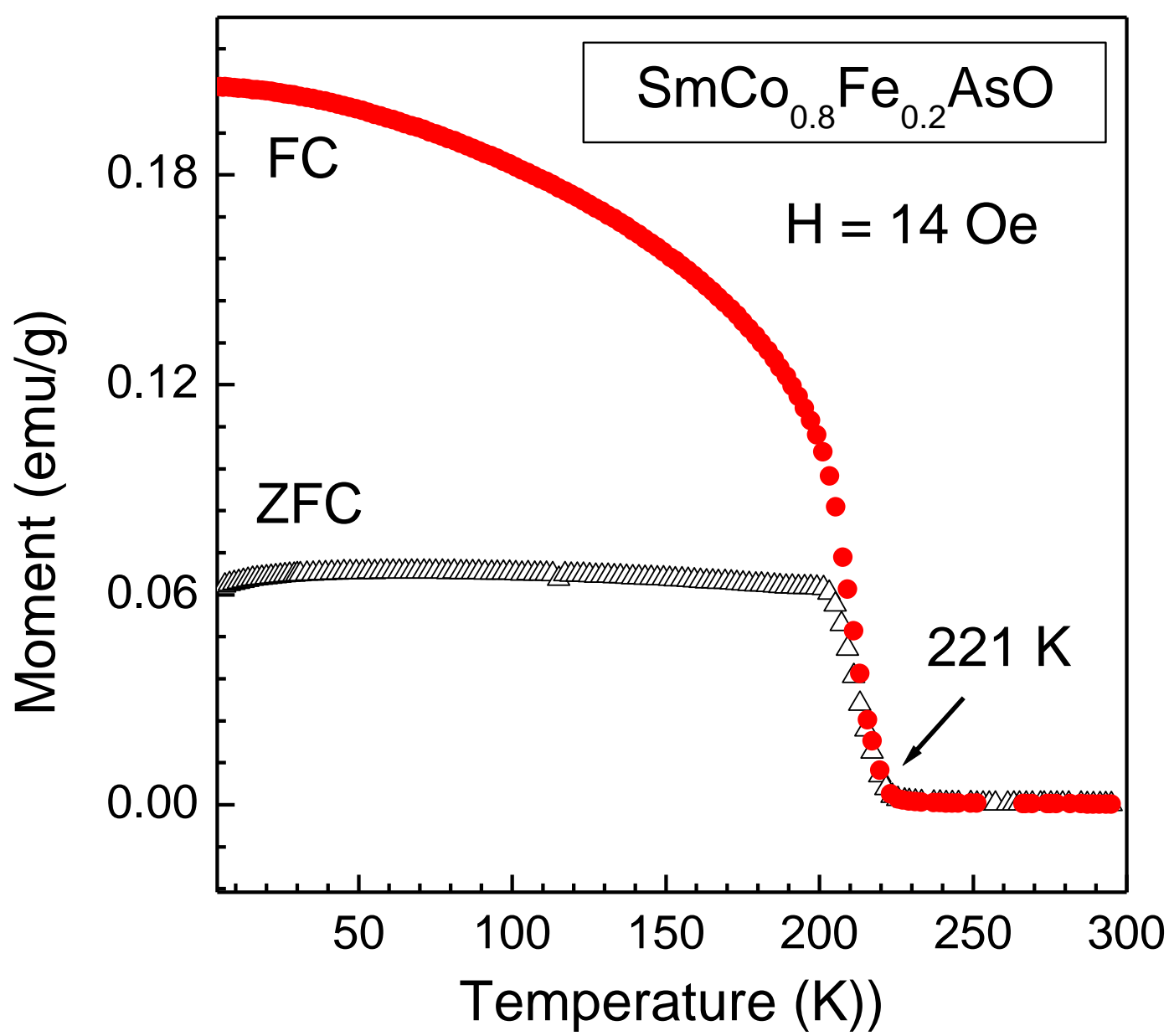

Figure 5: ZFC and FC branches of $SmCo_{0.8}Fe_{0.2}AsO$ measured at 14 Oe.

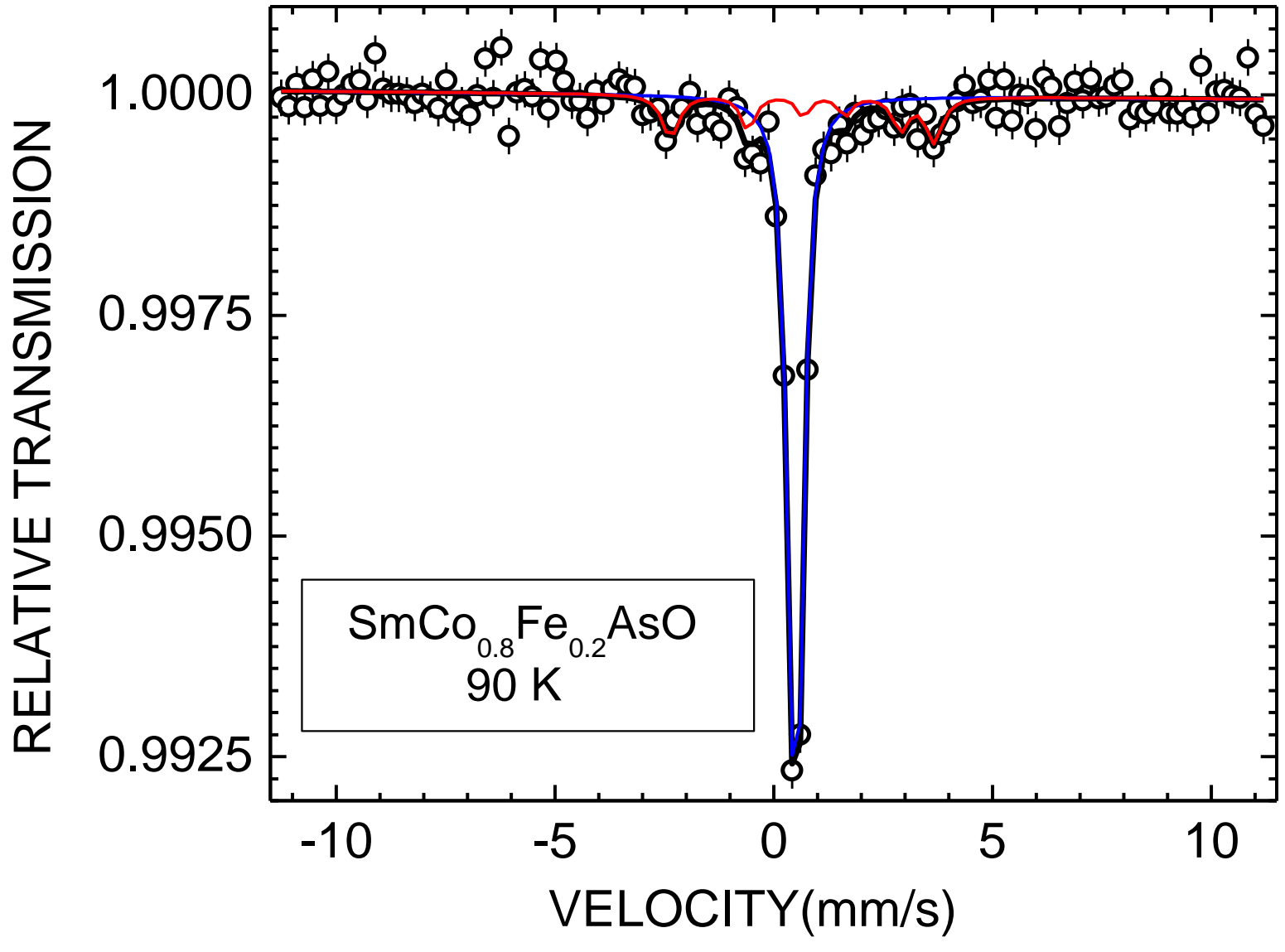

Figure 6: $^{57}$Fe Mossbauer spectrum of $SmCo_{0.8}Fe_{0.2}AsO$ at 90 K. Note the extra sextet originated from an extra magnetic phase.